\documentclass[sigconf]{acmart}

\AtBeginDocument{%
  \providecommand\BibTeX{{%
    \normalfont B\kern-0.5em{\scshape i\kern-0.25em b}\kern-0.8em\TeX}}}

\setcopyright{ccbysa}
\copyrightyear{2023}
\acmYear{2023}
\acmDOI{} 
 
\acmConference[IMI Workshop'23]{Intelligent Music Interfaces: When Interactive Assistance and Augmentation Meet Musical Instruments}{March 12,
  2023}{Glasgow, UK}
\acmBooktitle{Intelligent Music Interfaces: When Interactive Assistance and Augmentation Meet Musical Instruments, March 12, 2023, Glasgow, UK}

\begin{document}

\title{hDesigner: Real-Time Haptic Feedback Pattern Designer}

\author{Snehesh Shrestha$^1$}
\email{snehesh@umd.edu}
\orcid{0000-0002-1234-157X}
\affiliation{
  \institution{University of Maryland College Park}
  \city{College Park}
  \state{MD}
  \country{USA}
  \postcode{20740}
}

\author{Ishan Tamrakar$^1$}
\email{itamraka@umd.edu}
\orcid{0000-0002-6890-4917}
\affiliation{
  \institution{University of Maryland College Park}
  \city{College Park}
  \state{MD}
  \country{USA}
  \postcode{20740}
}

\author{Cornelia Fermuller}
\email{fermulcm@umd.edu}
\orcid{0000-0003-2044-2386}
\affiliation{%
  \institution{University of Maryland College Park}
  \city{College Park}
  \state{MD}
  \country{USA}
}

\author{Yiannis Aloimonos}
\email{jyaloimo@umd.edu}
\orcid{0000-0002-8152-4281}
\affiliation{%
  \institution{University of Maryland College Park}
  \city{College Park}
  \state{MD}
  \country{USA}
}

\renewcommand{\shortauthors}{Shrestha and Tamrakar, et al.}

\begin{abstract}
Haptic sensing can provide a new dimension to enhance people's musical and cinematic experiences. However, designing a haptic pattern is neither intuitive nor trivial \cite{seifi2020novice,mailvaganam2015haptic,ternes2008designing}. Imagined haptic patterns tend to be different from experienced ones. As a result, researchers use simple step-curve patterns to create haptic stimuli \cite{tom2020haptic, ternes2008designing}. To this end, we designed and developed an intuitive haptic pattern designer that lets you rapidly prototype creative patterns. Our simple architecture, wireless connectivity, and easy-to-program communication protocol make it modular and easy to scale. In this demo, workshop participants can select from a library of haptic patterns and design new ones. They can feel the pattern as they make changes in the user interface. With this new workflow, researchers and artists can design and rapidly test haptic patterns for downstream tasks such as research experiments or create new musical and cinematic experiences. More details about the project is available  at \url{https://www.snehesh.com/hDesigner}
\\
\\
\footnotemark{Equal Contribution}
\end{abstract}

\begin{teaserfigure}
  \includegraphics[width=\textwidth]{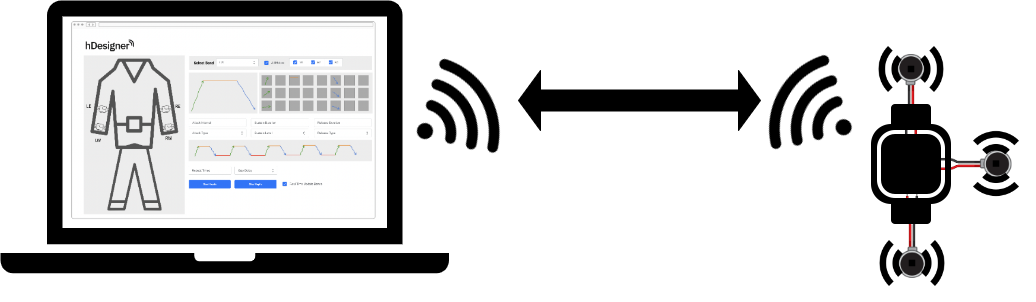}
  \caption{Haptic Designer User Interface wirelessly communicating with the haptic band with three motors.}
  \label{fig:teaser}
  \vspace{4mm}
\end{teaserfigure}

\maketitle

\section{Demo}
We will demo the haptic band and the hDesigner software at the workshop venue. Participants will wear the band, select from various presets, design their own custom patterns, and feel the vibration in real time with the band.

The user will wear the band at a desired location and turn them on. The device will connect to the server via a WiFi connection. Once hDesigner connects to the band, users can select an individual or group of motors they want to program. Then they can program the three segments of the pattern, i.e., the attack, the sustain, and the release (ASR). To program, they can either select from the presets that feature linear ramp-up and ramp-down, quadratic ease-in and ease-out, and square waves, each with different slopes. A more advanced level of control is also available where the users can specify the parameters of each segment to customize the preset or create their own new patterns. The advanced controls parameters are the minimum time step, the delta in milliseconds (ms), peak haptic intensity in percentage, minimum intensity, attack, and release curve types, and sustain duration.

To repeat the pattern, the user can specify the delay and the repeat values to complete the full pattern. The UI will update as changes are made in the design of any part of the haptic pattern. The user can also enable the real-time update to feel the haptic pattern as changes are made. The users can save the current presets and the entire pattern into the library to be reused later in the user's research or art projects.

The user can stop to interrupt the pattern at any time. They can also update to a new pattern at any time. During the demo, the user can experience various patterns such as a rhythmic beat, rotation effect, sliding effects, tapping effects, and heartbeats. The users can also test their ability to synchronize and tap to ranges of beats varied at different rates in beats per minute (BPM).

\section{\lowercase{h}Designer}
The overall setup is represented in figure \ref{fig:teaser}. The PC acts as the server, while each haptic band acts as the client. The PC server has three main modules:
\begin{itemize}
    \item The wireless communication module,
    \item A python server is a back end, and
    \item The web user interface (UI) is the system's front end.
\end{itemize}
The python server uses sockets to send packages to the client. It also generates Pulse Width Modulation (PWM) values based on the parameters transmitted from the user interface through an HTTP protocol. The user interface displays the ASR with each segment of the pattern with different colors. The UI also displays the overall pattern, including repetition. The UI shows a dedicated palette section for all the presets and the functionality to store new patterns in the library.

The messages from the server to the client are transmitted wirelessly via User Datagram Protocol (UDP) protocol. We chose UDP for its low latency transmission rate that minimizes communication delay. For each message, the receiving side sends a confirmation to the sender. If confirmation is not received within a configured time, the sender re-sends the message and retries up to 3 times. If no confirmation is received within the third try, the message is considered to have failed, and the user is notified.

On the client side, we have two main modules:
\begin{itemize}
    \item the wireless communication module, and
    \item the vibration generator module. The messages are in a type, length, and value (TLV) format.
\end{itemize}
These are sent to the client as a CSV string. The received packets are parsed, and vibration patterns are generated on the bands. The client also monitors for any interrupts to stop the current pattern or update to a new pattern at any time. For each message, the client responds with a confirmation message.

On the hardware side, the server runs on any computer that can connect to WiFi, run a web browser, and has python installed. The client is an ESP8266 with a WiFi chip and a Microcontroller with 8 General Purpose Input/Output (GPIO) pins capable of PWM. They are used to control three vibration motors independently. The circuitry is neatly integrated into a wearable band that is extremely light (200 gm), completely wireless, and battery-operated. The users can charge the thin and light LiPO batteries (20 gm) on or off the device.

\begin{figure}[ht]
    \centering
    \includegraphics[width=0.47\textwidth]{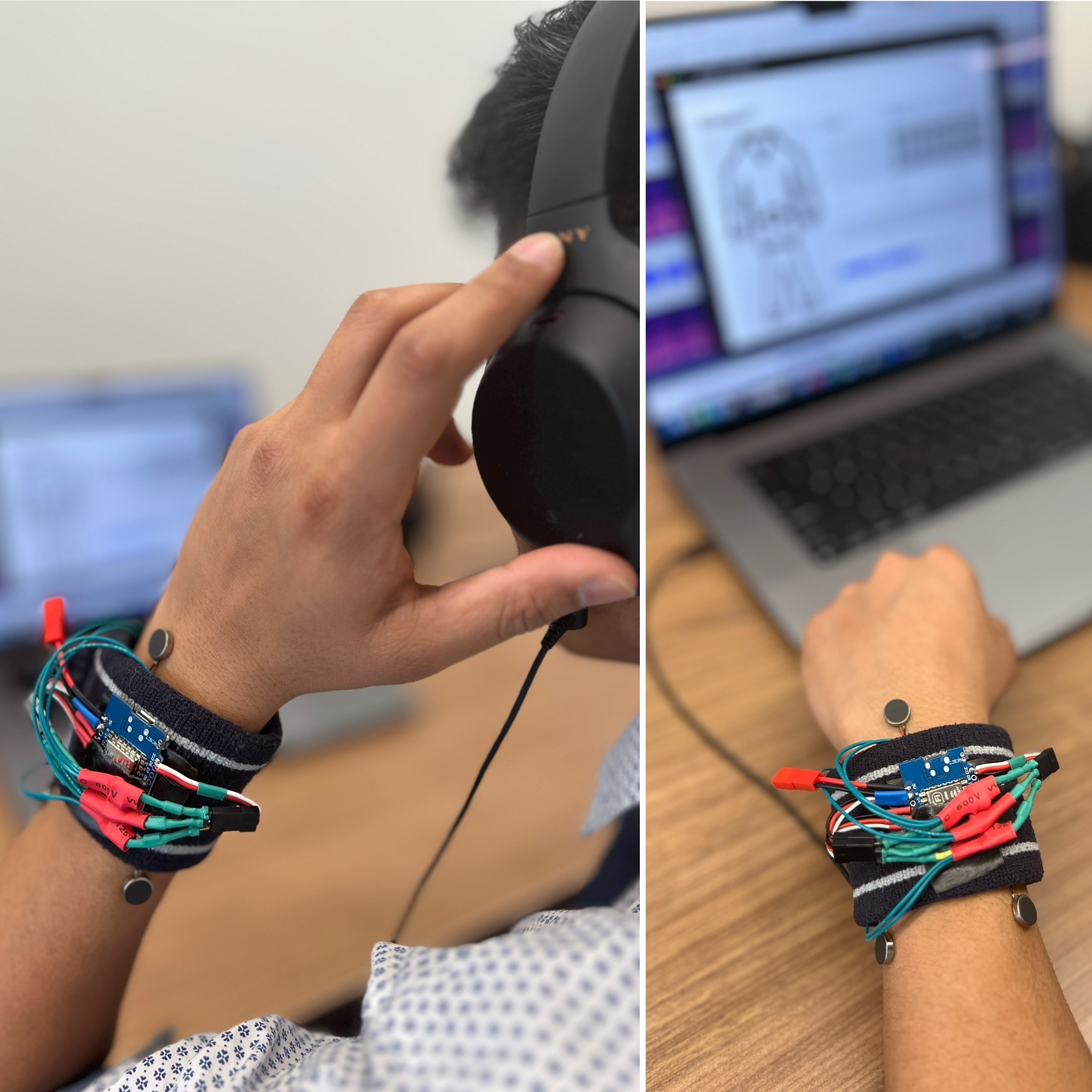}
    \caption{Illustration of the Haptic Band being used by a user}
    \label{fig:haptic_band}
\end{figure}

\section{Conclusion and Future Work}
In this interactive demo, users will have the opportunity to have a new kind of experience in a modality they are not accustomed to. We hope this experience will inspire them to design new creative experiences for music and movies. The hDesigner will also be released at the workshop, and the others can participate in contributing to the open-source project.

In the future, we plan a synchronization protocol and user interface that allows users to sync with media being played on their computer or smartphone. This will enable designers to integrate the haptic experience directly as they create and edit their music and videos.

\bibliographystyle{ACM-Reference-Format}
\bibliography{hDesigner}

\end{document}